\documentclass{cimento}

\usepackage{graphicx}  % got figures? uncomment this

\title{Flare forecasting and feature ranking using {\em{SDO/HMI}} data}
\author{M.~Piana\from{ins:x}\from{ins:z},
C. ~Campi\from{ins:y},
F.~Benvenuto\from{ins:x},
S.~Guastavino\from{ins:x} \atque
A. M.~Massone\from{ins:x}\from{ins:z}
}
\instlist{\inst{ins:x} Dipartimento di Matematica, Universit\`a di Genova, Genova, Italy
  \inst{ins:y} Dipartimento di Medicina, Universit\`a di Padova, Padova, Italy
  \inst{ins:z} CNR SPIN Genova, Genova, Italy}
\begin{document}

\maketitle

\begin{abstract}
We describe here the application of a machine learning method for flare forecasting using vectors of properties extracted from images provided by the {\em{Helioseismic and Magnetic Imager}} in the {\em{Solar Dynamics Observatory}} ({\em{SDO/HMI}}). We also discuss how the method can be used to quantitatively assess the impact of such properties on the prediction process.
\end{abstract}

\section{Introduction}
Solar flares, the most explosive phenomena in the heliosphere, may extend to over $10,000$ km while releasing more than $10^{32}$ erg in less than $100$ seconds, accelerating billions of tons of material to more than $10^6$ km/h, emitting electromagnetic radiation at all wavelengths and, in this way, triggering the whole space weather connection. The full comprehension of solar (and stellar) flare physics is still an open issue, to such an extent that we can talk about a sort of flare paradox: simple computation based on their physical and geometrical properties and on magnetohydrodynamic (MHD) equations would lead to predict a light-up time for flares longer than $10^5$ years, while the observed flash phase for these mysterious events is of the order of some minutes.

The numerical modelling of solar flare physics may rely on two different perspectives. On the one hand finite and boundary element methods applied against MHD partial differential equations allow the simulation of the electromagnetic fields and plasma properties in time and space; on the other hand, artificial intelligence allows pattern identification in the data mess and both source reconstruction with inverse methods and flare prediction with machine learning. In the present paper we will outline how a hybrid supervised/unsupervised machine learning method \cite{ref:beetal18} is able to utilize features extracted from experimental observations of active regions.

\section{Setup of the problem}
The data we consider in this paper are provided by the {\em{Helioseismic and Magnetic Imager}} in the payload of the {\em{Solar Dynamics Observatory}} ({\em{SDO/HMI}}). This telescope provides full disk vector magnetograms with a temporal cadence of $12$ minutes, starting from february 2010. Relying on the {\em{Solar Monitor Active Region Tracker (SMART)}} tools and also using pattern recognition methods developed within the 'Horizon 2020' FLARECAST effort (http://flarecast.eu), the input data at disposal for the machine learning analysis become feature vectors of dimension up to $171$ characterizing properties of the active regions (ARs) present in the {\em{SDO/HMI}} maps. Therefore the ingredients of our supervised approach are:
\begin{itemize}
\item A historical data set of feature vectors extracted from {\em{SDO/HMI}} data stored in the mission archive.
\item A set of labels, each one associated to a feature set and encoding the outcome information, i.e. labels testifying the possible flare occurrence and intensity.
\item A computational machine learning method that is trained by means of the historical (training) set and the corresponding set of labels.
\end{itemize}
When a new magnetogram arrives, the pattern recognition method extracts the features from it and the trained machine learning method both predicts the outcome corresponding to the new feature set and assesses the impact of each feature against the prediction effectiveness.

\section{The hybrid supervised/unsupervised machine learning method}
The method we used for both prediction and feature ranking \cite{ref:beetal18} is based on a hybrid supervised/unsupervised approach in which the supervised phase computes the weights with which each feature contributes to the prediction and the unsupervised phase provides an automatic, data-dependent threshold as a support for the binary prediction. More formally, we denote with $X$ the set of $N$ $F-$dimensional vectors representing the feature vectors at disposal in the training set. If $y$ is the $N-$dimensional vector containing the observed binary labels, the supervised step is realized by means of the LASSO approach \cite{ref:ti96}, which requires the solution of the minimum problem
\begin{equation}\label{eq:a}
{\hat{\beta}} = \arg \min_{\beta} (\|y - X \beta\|_2^2 + \lambda \| \beta \|_1)~.
\end{equation}
The solution of this problem is the vector ${\hat{\beta}}$ of feature weights, with which we can forward-compute theoretical labels by means of
\begin{equation}\label{b}
{\hat{y}} = X {\hat{\beta}}~.
\end{equation}
Then, we applied a fuzzy clustering technique \cite{ref:be81} on the computed labels in order to determine the centroids of the label-1 set (the flare occurs) and of the label-0 set (the flare does not occur). Therefore, when the new feature vector $x_{new}$ arrives, $x^{t}_{new} {\hat{\beta}}$ is compared to the two centroids in order to decide the cluster to which it belongs. Further, the entries of ${\hat{\beta}}$ quantify the impact of each feature on the prediction task; therefore ranking them corresponds to rank the prediction impacts. 

\section{Results}
\begin{figure}
\includegraphics[width=13.cm]{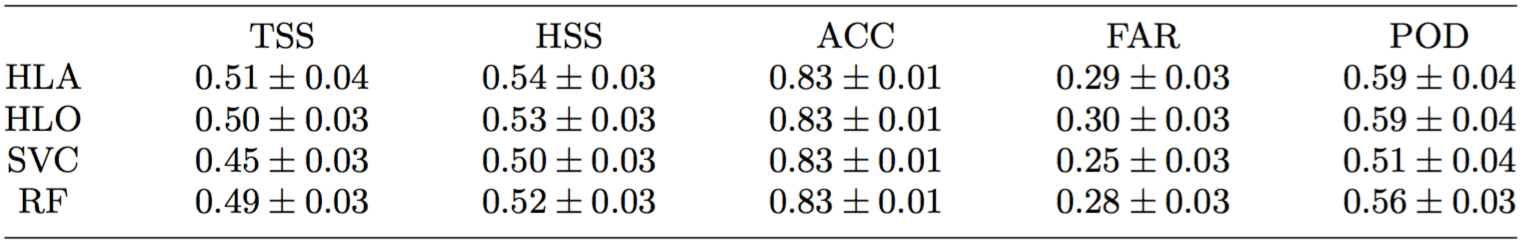}     % includes figure foo.eps
\caption{Skill scores obtained by the hybrid LASSO method described in this paper (HLA), a hybrid version of logit (HLO), a support vector machine for classification (SVC) and random forest (RF). TSS stands for true skill statistics, HSS for Heidke skill score, ACC for accuracy, FAR for false alarm ratio, POD for probability of detection. The numbers correspond to the average values over $100$ realizations of the training set with the corresponding standard deviations.}\label{fig:skill-scores}
\end{figure}

\begin{figure}
\includegraphics[width=13.cm]{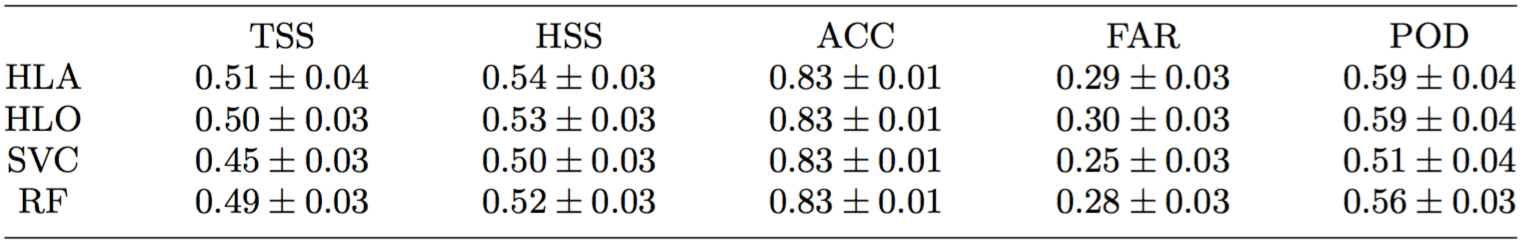}     % includes figure foo.eps
\caption{Ranking of the properties extracted from ARs recognized in {\em{SDO/HMI}} magnetograms as provided by the same machine learning methods considered in Table \ref{tab:skill-scores}. The numbers in the first four columns correspond to rank values computed from the weight vectors (\ref{eq:a}) and averaged over $100$ random realizations of the test set. The last two columns contain the average of the previous numbers over the four algorithm outcomes and the corresponding standard deviation, respectively.}\label{fig:ranking}
\end{figure}

We assessed the effectiveness of this approach for flare prediction and feature ranking by considering point-in-time {\em{SDO/HMI}} images in the time range between 09/14/2012 and 04/30/2016, with a time cadence of 24 hours. After the application of the pattern recognition step we had at disposal $4442$ sets of $171-$dimension feature vectors. Then the training set was realized by randomly extracting $66\%$ of the overall set of feature vectors and labeling each vector by annotating whether a flare with class at least $C1$ occurred in the next $24$ hours (the same can be done for flares with class at least $M1$). The set of the remaining feature vectors were used as test set for the experiment. Figure \ref{fig:skill-scores} explains the reliability of the prediction by comparing five standard skill scores \cite{ref:bletal12} computed by means of the hybrid method and of other three machine learning methods \cite{ref:wuetal09}, \cite{ref:cova95}, \cite{ref:br01}. Then, Figure \ref{fig:ranking} shows how these same methods are able to rank the features according to their impact on the prediction.

\section{Conclusions}
We showed how machine learning can be used for both binary flare forecasting and property ranking in the case of vector magnetograms provided by {\em{SDO/HMI}}. In particular, the ability to quantitatively assess the impact with which each property contributes to the prediction procedure has both physical and instrumental applications: high rank features are most likely associated to crucial physical processes and instruments that accurately observe properties with high rank are probably more worthwhile designing. 

The hybrid method described in this paper is currently implemented in the platform developed within the 'Horizon 2020' FLARECAST effort, together with other $15$ both supervised and unsupervised algorithms. This service is accessible for open download at http://flarecast.eu and is constantly updated also thanks to its highly modular design.

%\section{Examples}
%
%%\subsection{Tables}
%%Tables~\ref{tab:pricesI}, \ref{tab:pricesII} and~\ref{tab:pricesIII}
%%inserted at this point.

%\subsection{Mathematics}
%Here is a lettered array~(\ref{e.all}), with eqs.~(\ref{e.house})
%and~(\ref{e.phi}):
%\begin{eqnletter}
% \label{e.all}
% \drm x_\sy{F} & = & 1.2\cdot10^3\un{cm}, \qquad
%                     \tx{where\ } \sy{F} = \tx{Fermi}    \label{e.house}\\
% \phi_i        & = & i\pi                                \label{e.phi}
%\end{eqnletter}

%\subsection{Citations}
%We're almost done, just some citations~\cite{ref:apo}
%and we will be over~\cite{ref:pul,ref:bra}.
% 

\acknowledgments
The authors have been supported by the H2020 grant Flare Likelihood And Region Eruption foreCASTing (FLARECAST), project number 640216.

\end{document}